\documentclass[]{aa}  

\usepackage{graphicx}
\usepackage{hyperref}
\usepackage{txfonts}
\begin{document} 

   \title{APOGEE strings:  A fossil record of the gas kinematic structure}

   \author{A. Hacar
          \inst{1}
          \and
          J. Alves\inst{1}
          \and
          J. Forbrich\inst{1}
          \and
          S. Meingast \inst{1}
          \and
          K. Kubiak \inst{1}
          \and
          J. Gro\ss schedl \inst{1}
          }

   \institute{Department of Astrophysics, University of Vienna,
              T\"urkenschanzstrasse 17, A-1180 Vienna\\
              \email{alvaro.hacar@univie.ac.at}
             }

   \date{XXXX}

  \abstract
{We compare APOGEE radial velocities (RVs) of young stars in the Orion A cloud with CO line gas emission and find a correlation between the two at large scales  in agreement with previous studies. However, at smaller scales we find evidence for the presence of a substructure in the stellar velocity field. Using a friends-of-friends approach we identify 37 stellar groups with almost identical RVs. These groups are not randomly distributed, but form elongated chains or strings of stars with five or more members with low velocity dispersion across lengths of 1-1.5~pc.
The similarity between the kinematic properties of the APOGEE strings and the internal velocity field of the chains of dense cores and fibers recently identified in the dense interstellar medium is striking and suggests that for most of the Orion A cloud, young stars keep memory of the parental gas substructure where they originated.}

   \maketitle
%

\section{Introduction}\label{sec:intro}

Understanding the origin of the stellar motions inside molecular clouds is the final step in solving  the star formation puzzle. Solar-like stars are produced after the gravitational collapse of the gas inside dense cores  \citep[see][for a recent reviews]{BER07,AND14}. The subsequent connection between the gas and stellar velocity fields is a benchmark for the different star formation theories \citep[e.g.,][]{BAT03}. However, the simultaneous characterization of the gas and stellar motions inside clouds remains poorly studied in observations. 

From the comparison of the stellar radial velocities (RVs) and the motions of the dense cores in the \object{NGC~1333} protocluster in Perseus, \citet{FOS15} concluded that velocity dispersion of the young stars ($\sigma(V_{\star})=0.92\pm 0.12$~km~s$^{-1}$) is approximately twice as large as the velocity dispersion of the total core population within this region ($\sigma(V_{gas})=0.51\pm 0.05$~km~s$^{-1}$). Foster et al. have explained this mismatch as part of the global collapse of the NGC~1333 protocluster, a process that would produce a population of stars dynamically hotter than their parental cores.
Alternatively, these authors have proposed that the differences might be enhanced by the initial configuration of the dense gas that originally harbored these objects. 

In this paper we  explore the connection between the stellar velocity field and the internal substructure of the gas in the \object{Orion A} cloud \citep[D=414~pc;][]{MEN07} as a prototypical region in star formation studies.  
Both the young stellar population and gas properties within the Orion A cloud have been extensively characterized in the past using radio lines \citep{BAL87,NAG98,NIS15}, extinction \citep{LOM11}, far-IR continuum \citep{POL13,LOM14}, IR \citep{MEG12,MEG15,MEI16}, optical \citep{FUZ08,TOB09,BOU14}, and X-ray \citep{PIL13} observations.
In this work, we will combine these surveys with the latest stellar RV measurements provided by the new APO Galactic Evolution Experiment (APOGEE) observations in this cloud.


\section{APOGEE data: Stellar radial velocities in the Orion A cloud}\label{sec:data}

We used archival  APOGEE multiobject spectrograph \citep{WIL10} radial velocity data for the Orion A cloud, available at \url{www.sdss3.org}. Details on the data reduction and derivation of RVs are available at SDSS3 (DR12) web pages. The data used in this paper are part of the APOGEE ancillary science INfrared Spectroscopy of Young Nebulous Clusters (IN-SYNC) program \citep[]{COT14,FOS15}(see also \url{http://www.astro.ufl.edu/insync/}). A full description of this program, including the selection criteria and completeness of the IN-SYNC survey in \object{Orion}, are described in \citet{RIO15}.

A total of 2607 independent potential sources (EMBEDDEDCLUSTER\_STAR) were reported as part of the APOGEE survey in Orion A. Multi-epoch observations of different sources and individual fits to the data are listed as distinct entries in this catalog. The final radial velocity of each source is obtained by the mean of all the RVs available weighted by the number of visits  (NVISTS). 
A total of 1525 ($\sim~58\%$) matches are found between the APOGEE stars and the Spitzer YSOs (classified as protostars or stars with disk) detected by \citet{MEG12} in \object{Orion A} (937 individual matches; D~$\le$~1 arcsec) and/or the XMM point sources reported by \citet{PIL13} in \object{L1641} (979 individual matches at D~$\le$~3 arcsec, 391 of them also detected in Spitzer). Hereafter, we  refer to these subsamples as APOGEE-YSO and APOGEE-XMM sources, respectively. 
These two subsamples characterize the youngest stellar population of the Orion A cloud surveyed by APOGEE, i.e., from the earliest Class 0/I objects to the more evolved Class II/III objects.
Conversely, those APOGEE sources with no counterpart in any of the above Spitzer or XMM surveys are identified as part of the APOGEE-Field subsample (1082 sources).
In order to compare them to the gas velocities, the APOGEE synthetic heliocentric velocities (SYNTHVHELIO\_AVG, an average of the individual measured RVs using spectra cross-correlations with single best-match synthetic spectrum) were converted into LSR velocities (V$_{LSR}$) using the latest Galactic parameters described by \citet{REI09}. To ensure the quality of the RVs measurements, only sources with $\delta V_{LSR}\le$~0.25~km~s$^{-1}$ are considered in this study (i.e., 2368 or 90~\% of the total sample, including 887 APOGEE-YSO, 909 APOGEE-XMM, and 950 APOGEE-Field sources). 


\section{Results}

\subsection{Gas and stellar velocity fields: general properties}\label{sec:largescale}

   \begin{figure*}[!ht]
   \centering
   \includegraphics[width=0.9\linewidth]{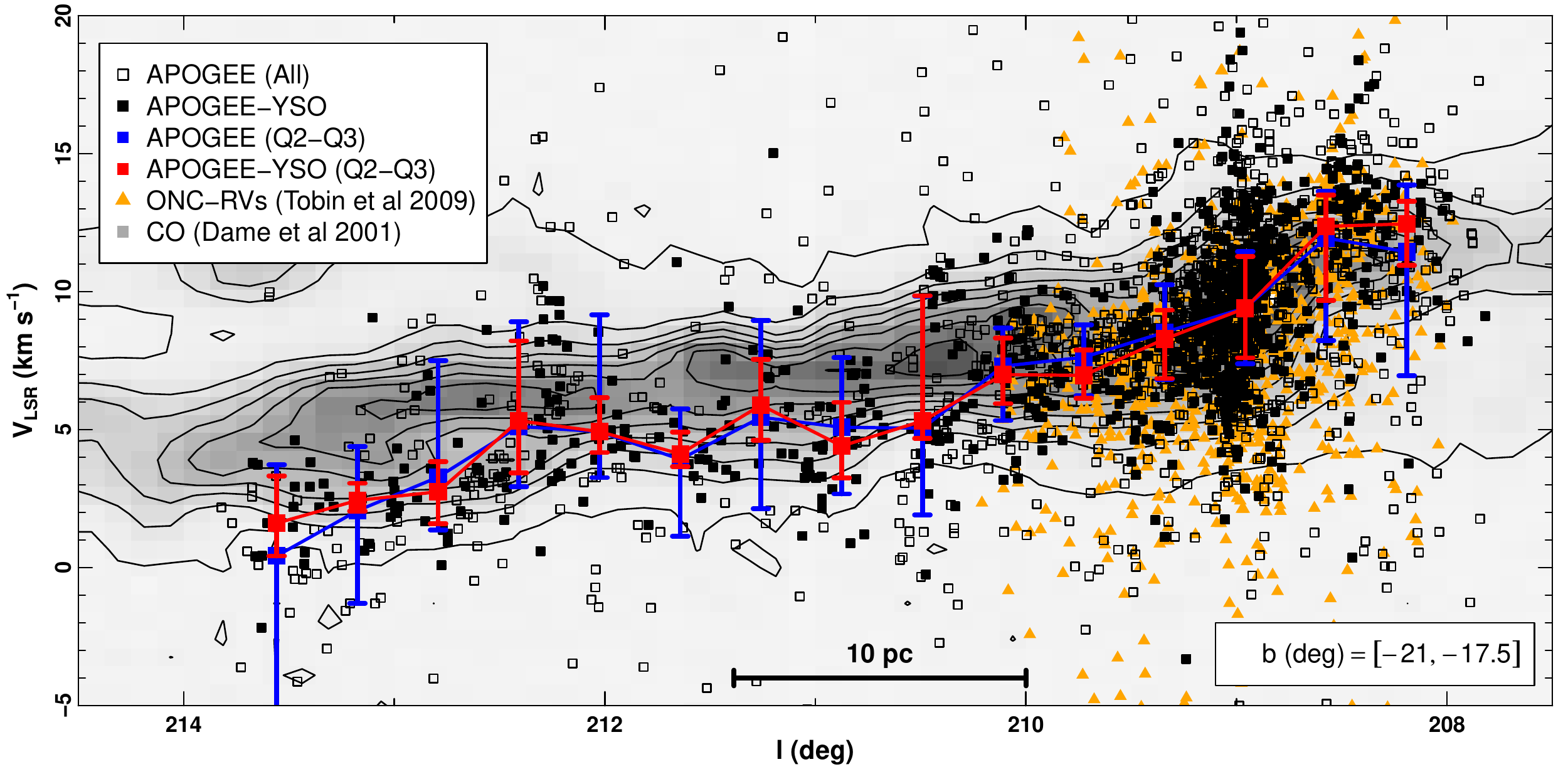}
   \caption{Position-velocity (PV) of both stellar RVs and gas emission along the Orion A cloud: (black empty squares) APOGEE sources, (black solid squares) APOGEE-YSOs, (orange triangles) sources detected by \citet{TOB09}, and (background; gray scale) average $^{12}$CO emission within Galactic latitudes -21.0~$\le b (deg) \le$~-17.5 \citep{DAM01}. The median (solid squares) and the 25-75\% (Q2-Q3) inter-quartile ranges (bars) for  all APOGEE sources and for the APOGEE-YSOs between Galactic longitudes 213.75~$\le l (deg) \le$~208 binned every 0.38~deg are represented in blue and red, respectively. The contours in the CO emission are spaced every 0.5 K~km~s$^{-1}$ starting at 0.1~K~km~s$^{-1}$.}
              \label{fig:PV_cut}%
    \end{figure*}

Figure~\ref{fig:PV_cut} displays the RVs of the APOGEE targets observed along the Orion A cloud as a function of the Galactic longitude $l$ (black squares). While the APOGEE sources are distributed along the whole velocity range, 80~\% of them are concentrated within a velocity range of V$_{LSR}=[0.0,15.5]$~km~s$^{-1}$. 
Binned in ranges of 0.38 deg in Galactic latitudes, both the mean and median values of the APOGEE RVs show a continuous velocity shift describing a global velocity gradient of 0.3 km~s$^{-1}$~pc$^{-1}$. Within these bins, the average 25-75\% (Q2-Q3) inter-quartile range presents a mean value of 5.6 km~s$^{-1}$ (blue bars). The average inter-quartile velocity difference falls  to 2.8 km~s$^{-1}$ when only these YSOs are included (red bars). Particularly in the case of the APOGEE-YSO subsample, both the mean and the inter-quartile ranges closely follow the gas emission traced in CO \citep{DAM01} from the northern Integral Shape Filament ($l\sim 208.5$ deg), throughout the Orion Nebula Cluster (\object{ONC}) ($l\sim 208.9$ deg), and towards the southernmost objects identified in the tail of the Orion A cloud ($l\ge 210$ deg). This correlation is even more striking when the binning effects (e.g., emission dilution, averages over large areas, and the asymmetric stellar distribution) are considered in the interpretation of this figure explaining the apparent shift between the APOGEE (i.e., its median and inter-quartile ranges) and gas velocities. 
In the particular case of the APOGEE-YSOs, a direct comparison between the stellar RVs and the CO channel maps shows how more than 80\% of the young stellar objects are found at the same velocities as the gas emission peaks in the Orion A cloud (see Sect.~\ref{ap:channelmaps}).

From their study of the stellar RVs, \citet{TOB09} found a close correlation between the stars in the ONC (orange triangles in Fig.~\ref{fig:PV_cut}) and the velocity structure of the $^{13}$CO emission within this region \citep{BAL87}. The comparison of the RVs obtained in the APOGEE sample with the large-scale $^{12}$CO observations \citep{DAM01,NIS15} supports these conclusions within those regions simultaneously covered in both surveys (i.e., 208~$\lesssim l \lesssim$~210; see also \citet{RIO15}). In addition, the larger coverage of the APOGEE survey allows us to extend these results to the entire Orion A cloud (i.e., 208~$\lesssim l \lesssim$~214). 

Moreover, and despite their larger velocity dispersions denoted by the presence of different spikes in the PV plot of Fig.~\ref{fig:PV_cut}, no significant variations are identified in the inter-quartile range of the RVs for the stars within clusters like the \object{ONC} (l~$\sim$~209~deg), \object{NGC~1977} (l~$\sim$~208.5~deg), \object{V380} (l~$\sim$~210.4~deg), or \object{L1641-S} (l~$\sim$~212.5~deg) in comparison to the rest of the cloud. Although beyond the scope of this paper, these results might suggest the presence of dynamically cold cluster halos with hot cluster cores characteristic of young stellar systems \citep[e.g.,][]{POR10}.

\subsection{Spatial and velocity correlations between gas and young stars}\label{ap:channelmaps}

   \begin{figure*}[ht]
   \centering
   \includegraphics[width=\textwidth]{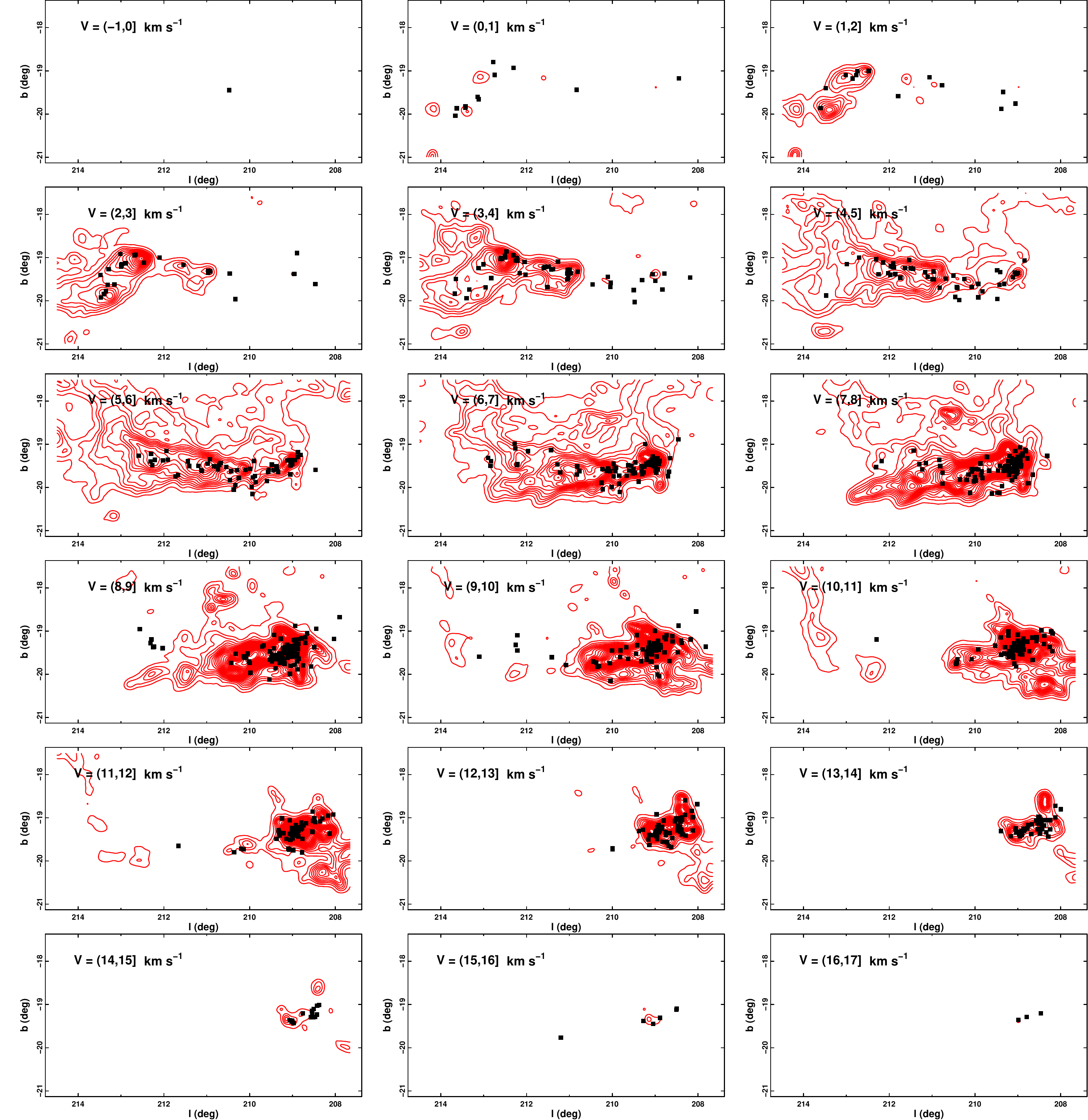}
   \caption{ Channel maps of the $^{12}$CO (2--1) emission in the Orion A cloud \citep[red contours;][]{NIS15} for LSR velocities between -1 and 17~km~s$^{-1}$ in channels of 1~km~s$^{-1}$ and contours every 0.6 K~km~s$^{-1}$. The maps cover Galactic latitudes between b~$=[-21,-17.5]$~deg, similar to those used to produce Fig.~\ref{fig:PV_cut}. Each channel map includes the position of the APOGEE-YSOs with RVs within the same velocity range (solid squares). There is a correlation between distribution of gas traced in CO and the young APOGEE sources with Spitzer counterparts both in space and velocity.}
              \label{fig:channelmaps}%
    \end{figure*}

The position-velocity (PV) diagram presented in Fig.~\ref{fig:PV_cut} shows the average properties of both gas and stellar velocity fields 
at different Galactic longitude bins and across 3.5~deg in Galactic latitude. 
However, the direct interpretation of this kind of PV plot is typically hampered 
by binning effects. 
Saturation, excitation, and dilution effects, as well as large-scale emission,  the intrinsic velocity variations inside the cloud and the evolutionary stage of the gas 
contribute to the average CO emission in velocity. On the other hand, the discrete position and reduced number of stars compared to the large areas
presenting CO emission only sample a limited range of the cloud motions. These comparisons are also affected by the different zero-point calibrations used in each case (e.g., by the use of different galactic parameters needed for the conversion between heliocentric and LSR velocities).
Differences in the mean and average values of the stellar and gas velocity fields might be apparent in these PV plots if large averages are considered, even when both components share the same intrinsic velocity structure.

To demonstrate these effects, in Fig.~\ref{fig:channelmaps} we show the distribution of the $^{12}$CO (2--1) emission in the entire Orion A cloud \citep{NIS15}
within the velocity range between -1 and 17~km~s$^{-1}$ in channels of 1~km~s$^{-1}$.
Although similar in extension to the previous $^{12}$CO (1--0) emission presented by \citet{DAM01},
these additional $^{12}$CO (2--1) data are preferred for this analysis because of their improved spectral and spatial resolutions. 
In addition to this CO emission, each of the channel maps includes the position of the APOGEE-YSOs with RVs falling into the same velocity range. 
 
As deduced from Figure~\ref{fig:channelmaps}, the APOGEE-YSOs are typically enclosed within the CO emission contours shown in these channel maps, equally spaced at integrated intensities of W$_{0.6}$=W($^{12}$CO (2--1))=~0.6~K~km~s$^{-1}$. In most of the cases, these sources are also located at the local emission peaks in their corresponding velocity maps.
The direct correlation between the CO motions and the APOGEE RVs reflects the intrinsic nature of the star formation process in clouds like Orion where stars and gas are closely related both spatially and in velocity during their early stages of evolution. 

   \begin{figure}[ht]
   \centering
   \includegraphics[width=\linewidth]{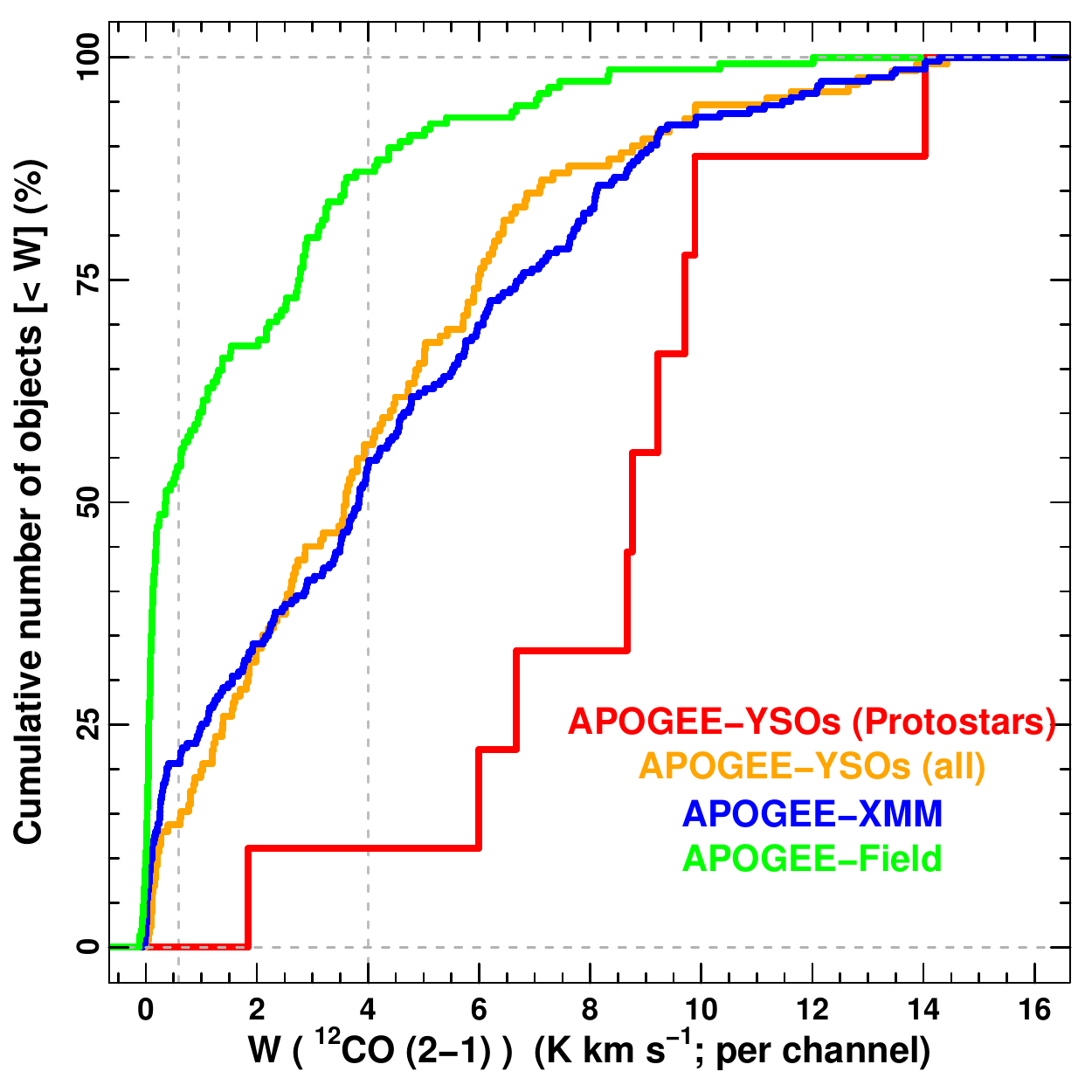}
   \caption{
   Cumulative function of APOGEE sources as a function of their corresponding local $^{12}$CO (2-1) emission at their corresponding velocity (see text for  description). Four subsamples are considered in this plot: APOGEE-YSO (orange), APOGEE-XMM (blue), APOGEE-Field (green), and APOGEE-YSOs (Protostars; red). The two emission thresholds considered in the text,  W$_{0.6}=$~0.6~K~km~s$^{-1}$ and W$_{4}=$~4.0~K~km~s$^{-1}$, are denoted by vertical dashed lines.
   }
              \label{fig:cumfunct}%
    \end{figure}

Additionally, Figure~\ref{fig:cumfunct} quantifies the time evolution of this correlation between stars and gas by presenting the cumulative fraction of objects at a given CO intensity within the channel maps shown in Fig.~\ref{fig:channelmaps} for Galactic latitudes l~$\ge$~210~deg. Four subsamples are considered in this diagram, namely, the APOGEE-YSO, the APOGEE-XMM, and the APOGEE-Field  stars already defined in Section~\ref{sec:data}, plus the APOGEE-YSO sources identified as protostars by \citet{MEG12}, hereafter referred to as APOGEE-YSOs (Protostars).  We  derived the W($^{12}$CO (2--1)) integrated emission of each APOGEE object as the mean CO emission at the position of the source within the 1~km~s$^{-1}$ channel map matching its RV. Approximately 85\% of the APOGEE-YSOs and 80\% of the APOGEE-XMM sources are found  to be associated with the CO emission levels of W$_{0.6}$ (see above). Following similar distributions at higher CO intensities, over~55\% of the objects in both samples are found in regions with W$_4$=W($^{12}$CO (2--1))$\ge$~4~K~km~s$^{-1}$. This correlation between stars and gas is maximized in the case of the more embedded APOGEE-YSO (Protostars) where $\sim$~80\% of these young sources are located inside regions with CO emission levels above W$_4$. Conversely, 52\% of the APOGEE-Field stars are found outside the first W$_{0.6}$ contour defined in our maps and only 13\% of them present emission levels higher than W$_4$.  
Although these field stars follow the kinematic structure of the cloud at large scales (Fig.~\ref{fig:PV_cut}), their distinct distribution is explained by the disruption of the gas around stars during their early evolution, diluting their local correlation with their gas envelopes at later stages. While no systematic difference can be found between the APOGEE-YSO and APOGEE-XMM populations (within the uncertainties), their comparison with the APOGEE-Field subsample suggests that young stars effectively dissipate their surrounding gas on a timescale comparable with the total evolution as  Class II/III objects \citep[cf.][]{STU16}.

Figure~\ref{fig:channelmaps} also illustrates how PV plots can be potentially biased by the large-scale emission of molecular tracers like CO. A large branch of CO gas devoid of stars is found at velocities between 6 and 9~km~s$^{-1}$ at Galactic coordinates l~$\sim [ 210-213 ]$~deg and b~$\sim$~-20.5~deg. Similarly, a large region presenting diffuse CO emission is found at l~$>$~213~deg and b~$>$~-19~deg at LSR velocities between 3 and 7~km~s$^{-1}$. These gas structures are redshifted by $\sim$~2-3~km~s$^{-1}$ with respect to the parallel, more compact star-forming structures of gas where the APOGEE-YSOs are found (compare channel maps between 4-5~km~s$^{-1}$ and 7-8~km~s$^{-1}$, respectively). When combined, the large-scale and uneven contributions of this additional CO emission explains the apparent shift between the mean CO velocities and the median and inter-quartile RVs of the APOGEE-YSO sample in the PV diagram presented in Fig.~\ref{fig:PV_cut}. While these PV plots are suitable for evaluating large-scale correlations (e.g., global trends, gradients), only detailed comparisons like those carried out in Fig.~\ref{fig:channelmaps} and Fig.~\ref{fig:cumfunct} can be used to evaluate the connection between stars and gas inside molecular clouds. 

\subsection{APOGEE strings: Identification and properties}\label{sec:strings}

   \begin{figure*}[!ht]
   \centering
   \includegraphics[width=0.95\textwidth]{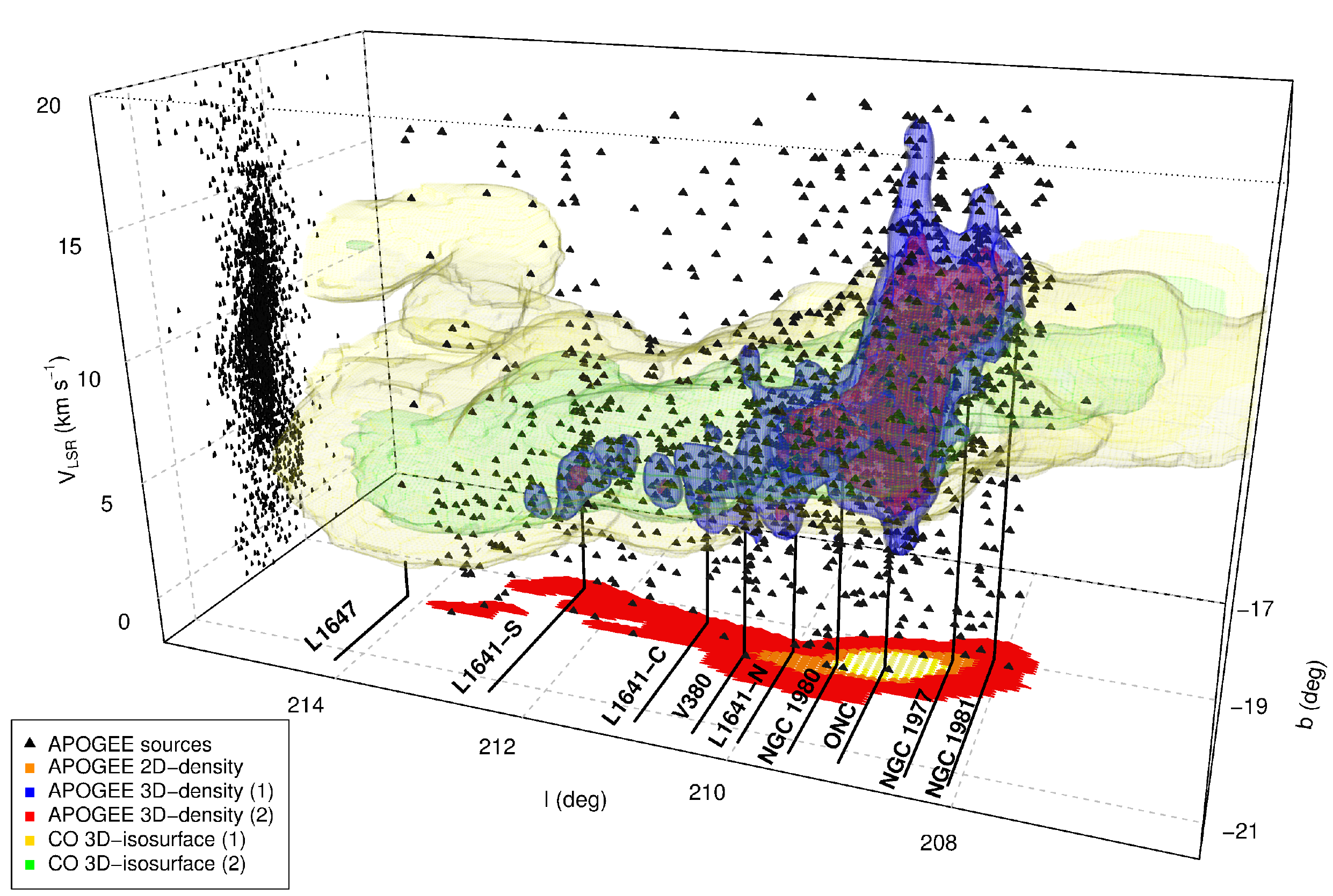}
   \caption{Position-position-velocity diagram (PPV) of the stellar and gas motions along the Orion A cloud: (black triangles) APOGEE 
stellar RVs; (blue and red surfaces) overdensities of APOGEE sources with 5 and 10 stars~pc$^{-2}$~(km~s$^{-1}$)$^{-1}$; (yellow and green) 0.5 and 2.0 K~km~s$^{-1}$ iso-contours of the $^{12}$CO emission \citep{DAM01} within the area surveyed by APOGEE. The PPV cube is spatially oriented with respect to the Galactic longitude ($b$; x-axis) and Galactic latitude ($l$; y-axis) coordinates, both in degrees units. Both gas and stellar velocities (V$_{LSR}$; z-axis)  refer to the LSR in km~s$^{-1}$ within the velocity range  V$_{LSR}=[-2.0,20.0]$~km~s$^{-1}$. The APOGEE sources are projected in the PP (xy) and PV (xz) planes. The spatial densities of $[$2,10,20,40) stars pc$^{-2}$ in the APOGEE sources is also color coded (from red to yellow) in the PP (xy) plane.  With the exception of the high stellar velocity dispersion found in clusters like the ONC or NGC~1977, 
we note how the stellar overdensities are typically clustered in the PPV space and nested within the velocity structure of the gas. The positions of the most important clusters within Orion A are indicated in the plot.
}
              \label{fig:PPV_cube}%
    \end{figure*}

Beyond the large-scale correlations presented in Fig.~\ref{fig:PV_cut}, an underlying kinematic substructure becomes apparent from the analysis of the stellar RVs surveyed by APOGEE. When plotted in the position-position-velocity (PPV) space, the observed APOGEE sources seem to cluster in different subgroups both spatially and in velocity. This property is highlighted in the PPV diagram presented in Figure~\ref{fig:PPV_cube}\footnote{
{An animation showing the rotation of this PPV cube is available at the following url: \url{https://sites.google.com/site/alvarohacarpersonalwebpage/projects/apogee_1} }
}. Using a 3D Gaussian kernel density estimate with a bandwidth of (0.1~deg, 0.1~deg, 0.2~km~s$^{-1}$), this figure displays the iso-contours defining the APOGEE overdensities of stars with 5 (blue) and 10 (red) stars~pc$^{-2}$~(km~s$^{-1}$)$^{-1}$. Stellar density values of 13, 15, and $>$~60 ~stars~pc$^{-2}$~(km~s$^{-1}$)$^{-1}$ 
are characteristic of highly populated clusters like NGC~1980, NGC~1981, and the ONC, respectively.
In addition, different and discrete stellar subgroups can be identified towards the tail of Orion A as local overdensities in the PPV space. The correspondence of these PPV overdensities with the position of the young clusters reported by \citet{MEG15} suggest a compact velocity configuration for these stellar groups, whose members share their kinematic properties with their neighboring companions.
 
The discrete nature of the RV measurements considered in the APOGEE survey suggests the use of a Friends-of-Friends approach \citep[FoF;][]{HUC82} as the most suitable and simplest way to identify the apparent stellar groups in the APOGEE data
simultaneously working in both spatial and velocity dimensions. The iterative procedure followed in this work can be summarized in a series of five steps. First, the procedure starts by selecting a reference source previously not assigned to any stellar group. Second,   we look for companions (friends) of this reference source within a projected distance L and with a velocity difference of $|V_{LSR,0}-V_{LSR,i}|\le \delta V_{\star,LSR}$. Third,  all the objects that fulfill these properties (if any) are added into a new group. Fourth,  each of these new members is taken as a reference for a new FoF search. Fifth, the association with a given group is finished when no additional source can be linked according to the above requirements. The above steps are repeated over the preselected APOGEE sample until no other source can be assigned to any additional new group. Those stars associated with groups with a number of members equal to or larger than N$_{members}$ are considered  stellar groups. On the contrary, those objects with fewer than N$_{members}$ companions in the PPV space are considered  isolated.  

   \begin{figure*}[ht]
   \centering
   \includegraphics[width=0.9\textwidth]{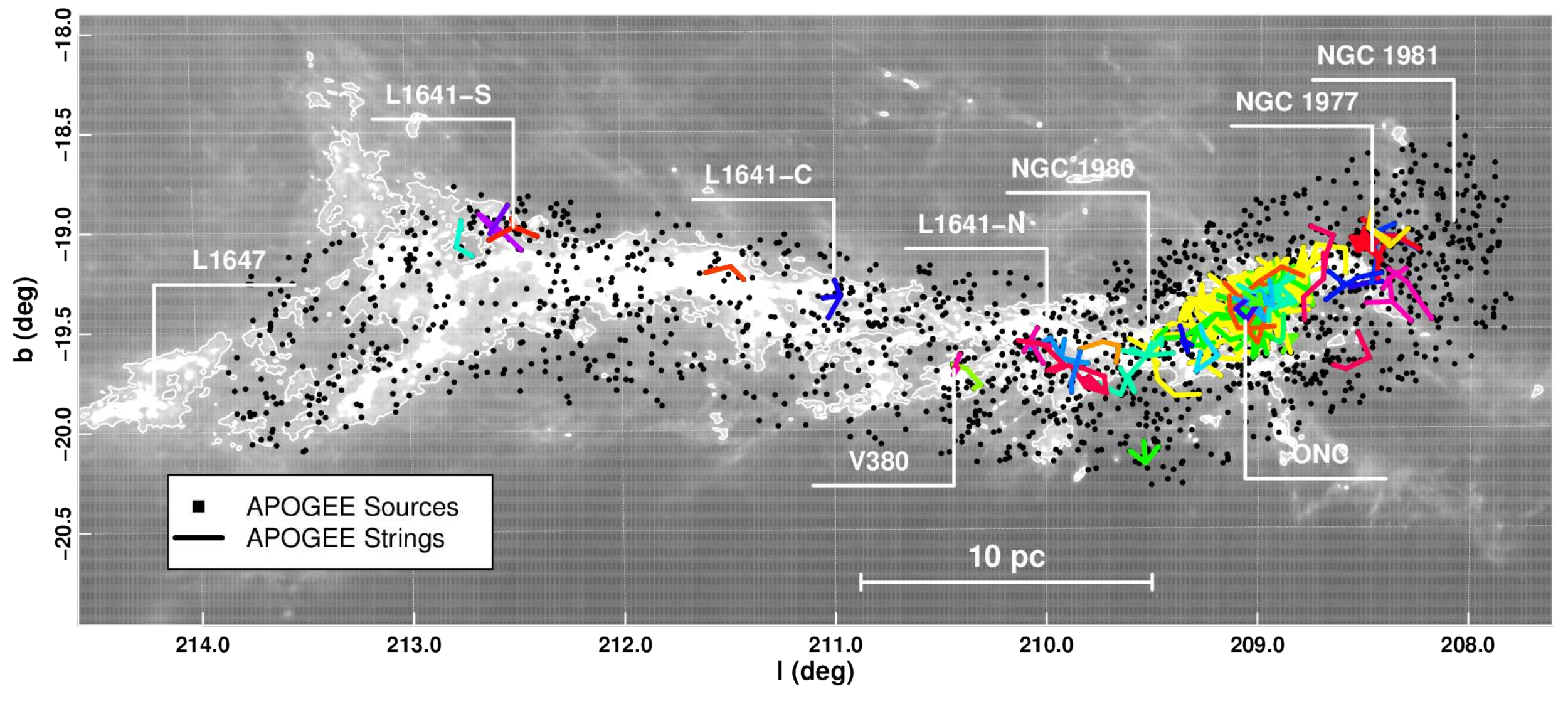}
   \caption{Spatial distribution of the APOGEE strings in the Orion A cloud: (Background) Herschel column density map \citep{LOM14}; (black points) APOGEE sources; (colored segments) APOGEE strings. The white contour encloses those regions with equivalent column densities of A$_V=$~5~mag. The positions of the most prominent clusters in the cloud are labeled in the figure.}
              \label{fig:OriA_strings}%
    \end{figure*}

Three initial parameters,  namely the linking length L, the velocity difference between stars $\delta V_{\star,LSR}$, and the number of members per string N$_{members}$, are required for the above FoF-based algorithm. Different tests indicate that the combination of the small velocity variations along the Orion A cloud and the high density of sources present in the APOGEE observations tend to favor the production of pairs and/or triplets (i.e., N$_{members}$=2 or 3; see Sect.~\ref{ap:contamination} for a discussion). To increase the significance of the results, we only consider large groups with at least N$_{members}\ge$~4 members. The velocity difference $\delta V_{\star,LSR}$  is assumed as the maximum error of the RVs measurements of the APOGEE pairs considered in each step, i.e., $max(\delta V_{\star,LSR}(0,i))\le 0.25$~km~s$^{-1}$. Finally, the linking distance L is then estimated from the maximum separation expected for two sources with a velocity difference equal to $max(\delta V_{\star,LSR})$ and an age younger than the total lifetime of the entire Orion A complex \citep[<~5~Myr;][]{BOU14}, i.e., L~$= 0.25$ km~s$^{-1} \times 5$~Myr~$ \sim$~1~pc. Although likely incomplete, this choice of parameters aims to identify those large stellar groups with almost indistinguishable kinematic properties. Stars in small groups (i.e., N$_{members}\le$~3), stars in groups with high velocity dispersion (i.e., $\delta V_{\star,LSR}>0.25$~km~s$^{-1}$) and/or in sparse associations (i.e., L~$>$~1~pc) would not be classified as group candidates.

   \begin{figure}[ht!]
   \centering
   \includegraphics[width=\linewidth]{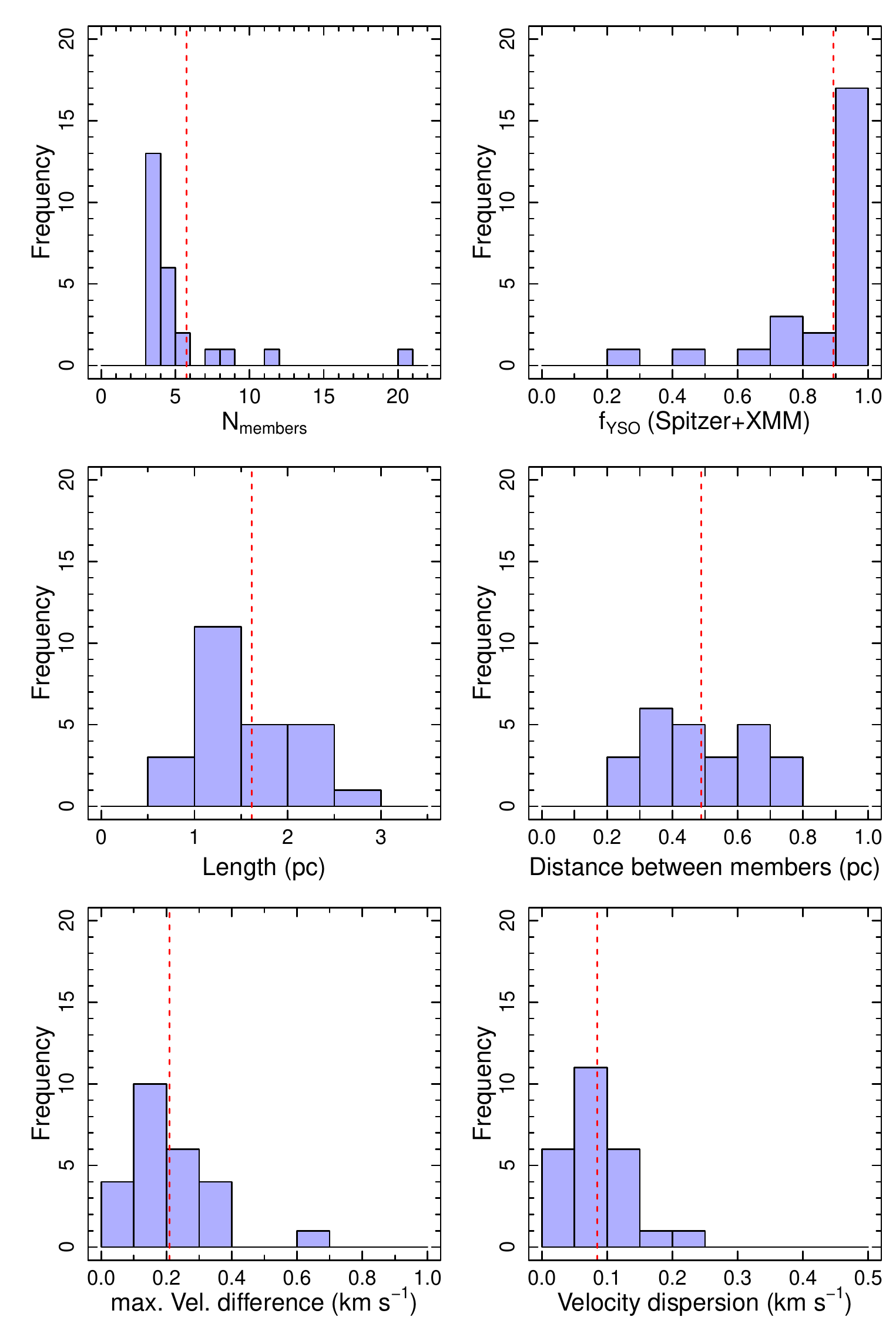}
   \caption{Statistical properties of the APOGEE strings
    (see also Table~\ref{tab:table1} available online).
     The blue histograms denote the normalized frequencies for the 25 APOGEE strings identified outside the Orion Nebula Cluster with positions l~$\ge$~209.3~deg or l~$\le$~208.7~deg. From left to right and from top to bottom: (a) Number of sources per string; (b) Fraction of young APOGEE objects (i.e., with Spitzer and/or XMM counterparts) per string; (c) Total length; (d) Mean distance to the nearest member of the string; (e) Maximum velocity difference within a string; (f) Internal velocity dispersion per string. The vertical red dashed lines indicate the mean value in each of the above distributions.
   }
              \label{fig:stats}%
    \end{figure}

According to the previous criteria, a total of 37 stellar groups are identified within the APOGEE survey along the Orion A cloud.  
The main properties of these 37 objects and their corresponding members are listed in Tables \ref{tab:table1} and \ref{tab:table2} available online.
Their spatial distribution is shown in Fig.~\ref{fig:OriA_strings}. The most numerous stellar group is identified at the \object{ONC} (with N=405 members). Additionally, 208 APOGEE sources are associated in 11 groups within the ONC region (208.7~$\le l~(deg)\le$~209.3), partially overlapping with both the NGC~1980 cluster and the OMC-2/3 ridge. Although this level of subclustering cannot be ruled out, the result illustrates the limitations and sensitivity of our simple algorithm that is only able to identify those groups of stars with almost indistinguishable kinematic properties. Excluding the ONC region, another 144 APOGEE sources are identified in 25 distinct groups by our FoF analysis. 
These last stellar groups appear to be organized in strings of stars, i.e., series of objects where the next member falls close to the direction defined by its two preceding companions. Based on this geometrical property, we have defined these stellar groups as APOGEE strings.

The statistical properties of the 25 APOGEE strings outside the ONC region are summarized in the histograms presented in Fig.~\ref{fig:stats}. On average, they present mean values of N$_{members}$=~5.7 members per group, sizes of 1.6~pc, a maximum velocity difference of $max(\mathrm{V}_{\star,LSR})-min(\mathrm{V}_{\star,LSR})=$~0.21~km~s$^{-1}$, and an average distance between nearest members of $\sim$~0.5~pc. 
Following the distribution of stars, most of these APOGEE strings are preferentially linked to the position of low-mass clusters like \object{L1641-N}, \object{V380,} or \object{L1641-S}. Approximately $58$~\% of the members in each of these chains of stars are identified as APOGEE-YSOs ($f_{YSO}$). This value increases  to 89\% if sources with X-ray emission are considered (i.e., APOGEE-XMM), 64\% of them (16 strings) which all their members present SPITZER and/or XMM counterparts.
Based on these $f_{YSO}$ ratios, and assuming a characteristic age of 2~Myrs for a Class II object \citep{EVA09}, we estimate a typical age of 2-3~Myrs for these strings.  While no a priori assumptions on any of these properties are made by our analysis technique, the stellar groups recovered in the APOGEE sample appear to correspond to elongated, parsec-size strings of relatively young stars.

\subsection{Statistical significance}\label{ap:contamination}

   \begin{figure}[!h]
   \centering
   \includegraphics[width=0.95\linewidth]{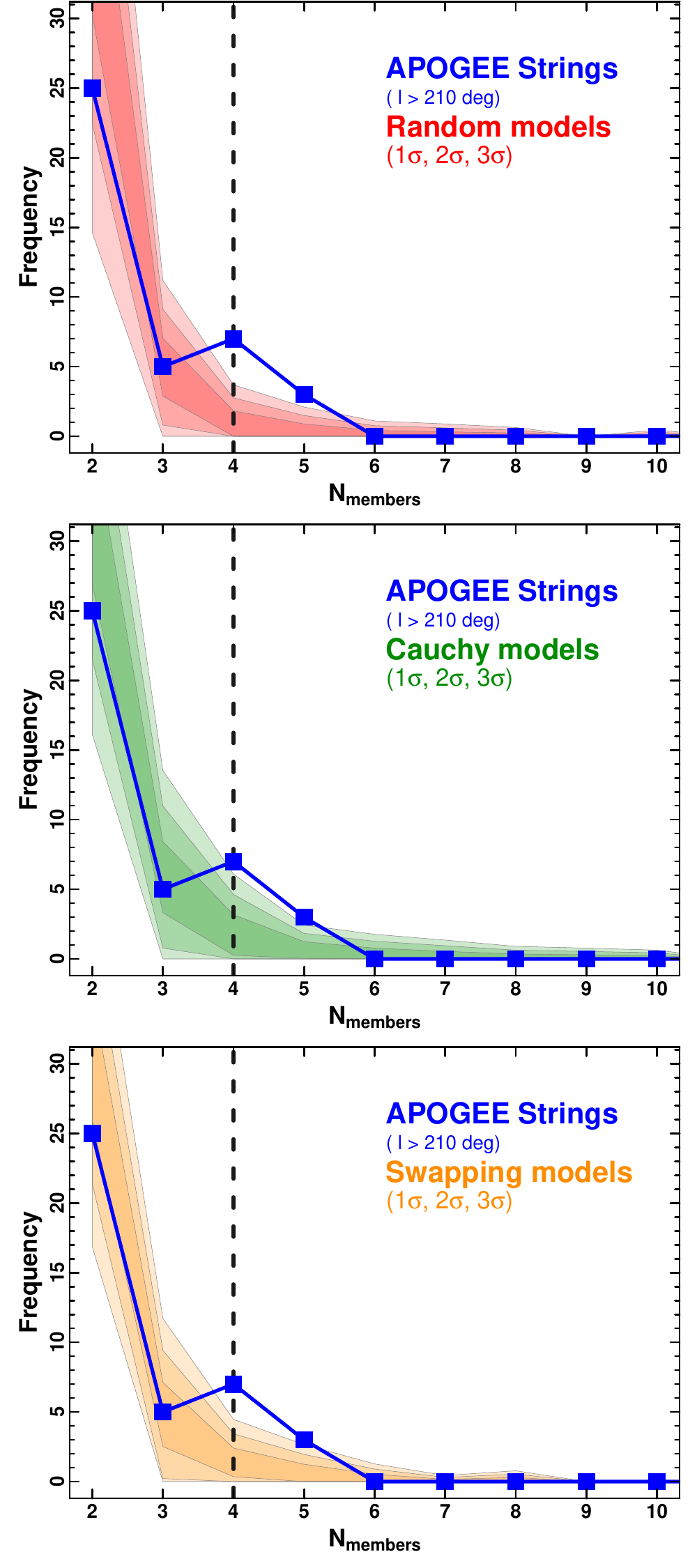}
   \caption{Frequency of strings detected in our Random (upper panel), Cauchy (mid panel), and Swapping (lower panel) models as functions of the number of members per string (N$_{members}$).
   The different shaded areas enclose the 1$\sigma$, 2$\sigma$, and 3$\sigma$ dispersion in each of these models. In all panels, the blue squares represent the detected APOGEE strings in Orion A at l~$\ge$~210~deg.
   The vertical dashed line indicates the minimum number of sources considered to define our APOGEE strings, with N$_{member}\ge$~4.
   }
              \label{fig:contamination}%
    \end{figure}

The use of FoF-based algorithms in highly populated datasets requires a careful characterization of their performance and the possible biases introduced by this analysis technique. In our particular case, the evaluation of the possible sources of contamination during the identification of the APOGEE strings (Sect.~\ref{sec:strings}) should take into account the intrinsic properties of the stellar velocity field and the observational constraints of the APOGEE data.   
First, this analysis should consider the incompleteness of the APOGEE sample targeting a small fraction ($<$~50\%) of the young stellar objects of identified in the Orion A cloud (see Sect.~\ref{sec:data}). By construction, this sample is systematically biased towards those objects with low extinctions owing to the sensitivity of these observations and the S/N and number of lines necessary to obtain their RVs \citep[see][for a discussion]{RIO15}. Second, the stellar field traced by this APOGEE sample contains a non-negligible contribution of outliers, binaries, and foreground and background sources randomly distributed along the sampled area. Third, any characterization of these effects should take into account the intrinsic clustering of the stellar population of the Orion A cloud \citep{MEG15} and the influence of these clusters in the observed stellar field (see the large velocity spreads created at the positions of clusters like the ONC at l~$\sim$~209~deg or L1641-S at l~$\sim$~212.2~deg in Figs.~\ref{fig:PV_cut} and \ref{fig:PPV_cube}).  

Owing to the above caveats, we have tested our FoF algorithm outside of the southern part of the Orion A cloud, i.e., within the region  l~$\le$~210~deg, containing a total number of 740 APOGEE sources (hereafter N$_{test}$).  This selection deliberately avoids highly populated clusters like the ONC, allowing us to directly evaluate the contamination of our analysis technique by random alignments. Within this area, we have investigated the performance of our FoF algorithm using the same linking parameters L~$\le$~1~pc and $\delta V_{\star,LSR}\le 0.25$~km~s$^{-1}$ as in Sect.~\ref{sec:strings} for groups of N$_{members}\ge$~2 objects in three representative cloud models, namely, the Random, Cauchy, and Swapping models (see below). We have quantified the performance of this algorithm by running each of these models 50 times. 
Figure~\ref{fig:contamination} shows the number of strings detected as a function of the group members (N$_{members}$) in these models in comparison with the 40 groups recovered within this region in the original APOGEE sample (10 of which have N$_{members}\ge$~4 members). For each N$_{members}$ value, we represent the 1$\sigma$, 2$\sigma$, and 3$\sigma$ variation ranges of these models over their mean value, estimated by multiplying their standard deviation $\sigma$ per bin by its corresponding factor.

First, we applied our algorithm to a sample of N$_{test}$ stars whose positions and velocities are randomized following a uniform probability density distribution within the full range of values detected in this region, l~$\in [210,213.9]$~deg, b~$\in [-20.3,-18.8]$~deg, and V$_{LSR}\in [0,10]$~km~s$^{-1}$. These Random models evaluate the less favorable case for our analysis where any detection can be attributed to a chance alignment of objects. As illustrated in Fig.~\ref{fig:contamination} (upper panel), although several random pairs and triples (N$_{members}$=~2 or 3) are recovered by our algorithm, the detection of random alignments rapidly decreases for groups with more than 3 objects. The $\gtrsim 5\sigma$ differences between these fully random models compared to our APOGEE strings with N$_{members}=$~4 \& 5 objects denote the robustness of our algorithm identifying strings at large N$_{members}$ values. 

A second set of models assumes a sample of N$_{test}$ stars at the same positions as our APOGEE sources whose RVs follow a Cauchy distribution in each of the bins defined in Fig.~\ref{fig:PV_cut}. The Cauchy distribution (also referred to as the Lorentz-Cauchy distribution) resembles a Gaussian-like distribution with extended non-Gaussian wings and is found to describe the spectral parameters derived from the APOGEE data \citep{COT14} as well as the local RVs per bin along the the Orion A cloud. 
In practice, these distributions can be defined by the number of objects per bin and the Q2-Q3 inter-quartile ranges introduced in Sect.~\ref{sec:largescale}. 
Compared to a full randomization, the Cauchy models preserve the spatial distribution of stars and the kinematic properties of the cloud at large scales (i.e., gradients and local dispersions).  Additionally, they allow us to test the correlation between the spatial and velocity distributions of stars within this cloud. 
Figure~\ref{fig:contamination} (mid panel) presents the number of strings identified in these models, again as function of N$_{member}$. The spatial correlation between sources and the tighter velocity structure of the cloud increases the probability of detecting large groups of stars by a factor of $\sim 2$ compared to the previous Random models. Still, the signals of the observed APOGEE strings with N$_{members}\ge$~4 members are found at significant values of $>3\sigma$ compared to those groups detected in the Cauchy models. 

Finally, and in a third set of models, we keep the positions of the APOGEE sources but swap their RVs among the sources within each of the Fig.~\ref{fig:PV_cut} bins. 
This last scenario is designed to statistically explore the quality and completeness of our observations. As shown in Fig.~\ref{fig:contamination} (lower panel), the number of groups recovered in these Swapping models closely reproduce the results obtained in our first Random models (upper panel). While these new models present the same spatial and velocity structures as the observed APOGEE data, their differences are explained by the sensitivity of the FoF algorithms to discontinuities along the strings (e.g., if similar RVs of the sources in a group are exchanged by those values of other stars with larger relative velocities causing the loss of this structure).

The results presented in Fig.~\ref{fig:contamination} confirm the statistical significance of the strings of stars detected in the APOGEE data. In all explored scenarios, the signal detected in the original APOGEE sample is always found at values of $>3\sigma$ compared to our simulations for groups with N$_{members}\ge$~4 members. Indeed, and in all the 150 simulations that were run for these tests, the number of strings recovered with N$_{members}=$~4 or 5 members in these models always remain  below the number of detections reported in our APOGEE data. Although the detection of smaller groups of stars cannot be ruled out, these comparisons also justify the selection criteria in our FoF algorithm with min(N$_{members}$)=~4 (see Sect.~\ref{sec:strings}). Combining the results of the three previous models we estimate an average contamination of 19~\% and 16~\% for these strings with N$_{members}=$~4 and 5 members, respectively.

\section{Discussion: Fibers, chains of cores, and the formation of strings of stars}\label{sec:fibers}

From the study of their spatial distribution, \citet{GOM93} concluded that the TTauri stars in the \object{Taurus} molecular cloud are clustered forming elongated groups with radii between 0.5-1.0~pc. These authors hypothesize that in order to be able to recognize such stellar groups their internal velocity dispersion should be $\le~0.5$~km~s$^{-1}$. Using large-scale millimeter line maps in the same cloud, \citet{TAF15} found that the dense cores identified within the \object{B213/L1495} filament are preferentially found forming chain-like groups of $\ge$~2~dense cores about $\sim$~0.5~pc long with internal velocity dispersion on the order of the sound speed or $\sigma(V_{gas})=$~0.2~km~s$^{-1}$. As demonstrated by \citet{HAC13}, these elongated groups of dense cores are generated during the quasi-static gravitational fragmentation of individual velocity coherent fibers. 

As part of the above fiber-to-core transition, the low internal velocity dispersion along the observed chains of cores are the result of the sonic internal velocity structure of their parental fibers. It is then expected that the low-mass stars formed after the collapse of these chains of cores would also present similarly low velocity differences. 
After the dissipation of their envelopes, these chains of cores would lead into similarly ordered strings of stars. Moving ballistically throughout the cloud, and particularly during the early stages of evolution, some of these strings would still be recognized as roughly aligned groups of stars with almost identical kinematic properties. 

The strong similarities of the APOGEE strings detected in Orion A with the expected properties of the stars created in the aforementioned chains of cores (i.e., number of members, internal velocity dispersion, and total length) suggest that these stellar subgroups might have originated as part of the original gas kinematic structure. This hypothesis is favored by the highly filamentary gas structure reported in this cloud, forming complex and intertwined networks of quiescent filaments \citep[e.g.,][]{NAG98,POL13} similar to those bundles of fibers identified in Taurus \citep{HAC13}. Thus, the detection of multiple APOGEE strings with a velocity dispersion on the order of the sound speed might represent a fossil record of the internal velocity structure of their parental fibers. This primordial velocity structure would still be recognizable in most of the low-mass clusters and extended population of stars in the Orion A cloud. Closely packed in the PPV space in regions like \object{L1641-N} or \object{V380}, compact associations of these APOGEE strings would appear as stellar groups with larger velocity dispersion. Conversely, and apparent in dense clusters like the ONC, two-body relaxation processes could have rapidly erased these characteristic kinematic signatures, as has been suggested by recent simulations \citep[e.g.,][]{KUZ15}. 

\section{Conclusions}
In this paper, we have investigated the internal substructure of the RVs of the stars included by the APOGEE survey in the Orion A cloud. The main results of this work are summarized as follows: 

   \begin{enumerate}
      \item Overall, the stellar velocity field traced by the RVs of the stars closely follows the large-scale velocity gradient of the gas in Orion A. Confirming previous studies in the vicinity of the ONC \citep{FUZ08,TOB09}, the combination of APOGEE data and wide-field CO maps allow us to extend this conclusion to the entire Orion A cloud.
      \item The direct comparison of the APOGEE RVs with the CO channel maps demonstrates that more than 80\% of the young APOGEE sources with Spitzer and/or XMM counterparts are directly associated with regions with high CO emission within this cloud, both spatially and in velocity. Conversely, the low occurrence of these correlations for those field stars surveyed by APOGEE suggests that the newborn stars effectively dissipate their surrounding gas during their Class II/III phase.
      \item In addition to these global properties, and for the first time, an analysis of the stellar densities in the PPV space reveals the presence of an underlying kinematic structure in the stellar velocity field. Using a FoF-based approach, we have identified 37 different groups of nearby stars with almost indistinguishable kinematic properties. Forming elongated chains, these APOGEE strings present characteristic lengths of $\sim$~1.5~pc, a mean number of five members, and a maximum velocity difference of 0.21~km~s$^{-1}$. 
      \item The strong similarities between the kinematic properties of these APOGEE strings and the internal velocity field of the chains of dense cores and fibers identified in Taurus suggest that these APOGEE strings might keep the memory of the parental gas substructure where they  originated.
   \end{enumerate}

\begin{acknowledgements}
      The authors thank Bob O'Dell for his constructive referee comments that helped to improve the content of this paper.
      Funding for SDSS-III has been provided by the Alfred P. Sloan Foundation, the Participating Institutions, the National Science Foundation, and the U.S. Department of Energy Office of Science. The SDSS-III web site is http://www.sdss3.org/.
SDSS-III is managed by the Astrophysical Research Consortium for the Participating Institutions of the SDSS-III Collaboration including the University of Arizona, the Brazilian Participation Group, Brookhaven National Laboratory, Carnegie Mellon University, University of Florida, the French Participation Group, the German Participation Group, Harvard University, the Instituto de Astrofisica de Canarias, the Michigan State/Notre Dame/JINA Participation Group, Johns Hopkins University, Lawrence Berkeley National Laboratory, Max Planck Institute for Astrophysics, Max Planck Institute for Extraterrestrial Physics, New Mexico State University, New York University, Ohio State University, Pennsylvania State University, University of Portsmouth, Princeton University, the Spanish Participation Group, University of Tokyo, University of Utah, Vanderbilt University, University of Virginia, University of Washington, and Yale University.
\end{acknowledgements}

\bibliographystyle{aa} 

\bibliography{APOGEE_strings}

\Online


\begin{table*} [!h]
\caption{APOGEE strings: Main properties}\label{tab:table1} 
\centering 
\begin{tabular}{l c c c c c c c c c} 
\hline 
ID &	$\langle l \rangle$ &		$\langle b \rangle$ &	N$_{members}$ &	f$_{YSO}^{(2)}$ &	L &	 $\delta\mathrm{V}_{max}$ &	$\sigma$V &	$\langle D_{near} \rangle$ &	Region  \\ 
&	(deg) &	(deg) &	&	&	(pc) &	(km s$^{-1}$) &	(km s$^{-1}$) &	(pc) \\  
\hline
 1 &	208.288 &	-19.334 &	  4 &	1.00 &	2.3 &	0.15 &	0.08 &	0.77 &	   NGC1981	\\ 
 2 &	208.381 &	-19.342 &	  6 &	0.67 &	1.9 &	0.09 &	0.04 &	0.64 &	   NGC1977	\\ 
 3 &	208.400 &	-19.110 &	  6 &	0.83 &	2.3 &	0.12 &	0.05 &	0.72 &	   NGC1977	\\ 
 4 &	208.411 &	-19.025 &	  5 &	0.80 &	1.5 &	0.11 &	0.05 &	0.50 &	   NGC1977	\\ 
 5 &	208.455 &	-19.080 &	 21 &	0.90 &	2.3 &	0.65 &	0.20 &	0.21 &	   NGC1977	\\ 
 6 &	208.512 &	-19.293 &	  5 &	0.80 &	1.1 &	0.15 &	0.07 &	0.25 &	   NGC1977	\\ 
 7 &	208.564 &	-19.657 &	  4 &	0.25 &	1.4 &	0.18 &	0.07 &	0.70 &	   NGC1977	\\ 
 8 &	208.586 &	-19.282 &	  5 &	1.00 &	2.1 &	0.23 &	0.09 &	0.69 &	   NGC1977	\\ 
 9 &	208.727 &	-19.209 &	  7 &	1.00 &	3.4 &	0.42 &	0.16 &	0.61 &	       ONC	\\ 
10 &	208.918 &	-19.536 &	  4 &	1.00 &	1.0 &	0.07 &	0.04 &	0.51 &	       ONC	\\ 
11 &	208.937 &	-19.356 &	 12 &	0.75 &	2.0 &	0.30 &	0.09 &	0.33 &	       ONC	\\ 
12 &	208.943 &	-19.356 &	 58 &	0.86 &	3.4 &	1.45 &	0.42 &	0.20 &	       ONC	\\ 
13 &	208.996 &	-19.356 &	  6 &	0.50 &	1.0 &	0.33 &	0.12 &	0.31 &	       ONC	\\ 
14 &	209.014 &	-19.409 &	 21 &	0.86 &	2.9 &	0.40 &	0.11 &	0.32 &	       ONC	\\ 
15 &	209.015 &	-19.416 &	  6 &	0.33 &	1.3 &	0.12 &	0.04 &	0.38 &	       ONC	\\ 
16 &	209.025 &	-19.401 &	  6 &	1.00 &	1.2 &	0.15 &	0.06 &	0.35 &	       ONC	\\ 
17 &	209.026 &	-19.374 &	  4 &	0.75 &	0.7 &	0.10 &	0.05 &	0.37 &	       ONC	\\ 
18 &	209.046 &	-19.407 &	405 &	0.83 &	8.1 &	4.20 &	1.12 &	0.10 &	       ONC	\\ 
19 &	209.060 &	-19.437 &	 79 &	0.81 &	5.4 &	1.16 &	0.33 &	0.21 &	       ONC	\\ 
20 &	209.261 &	-19.623 &	  5 &	1.00 &	1.7 &	0.14 &	0.06 &	0.53 &	       ONC	\\ 
21 &	209.348 &	-19.586 &	  4 &	1.00 &	1.0 &	0.22 &	0.10 &	0.36 &	   NGC1980	\\ 
22 &	209.539 &	-20.138 &	  4 &	0.50 &	1.0 &	0.21 &	0.10 &	0.59 &	   NGC1980	\\ 
23 &	209.574 &	-19.695 &	 12 &	0.83 &	2.8 &	0.32 &	0.10 &	0.46 &	   NGC1980	\\ 
24 &	209.729 &	-19.633 &	  4 &	0.75 &	1.3 &	0.35 &	0.14 &	0.65 &	   L1641-N	\\ 
25 &	209.826 &	-19.763 &	  9 &	1.00 &	2.0 &	0.39 &	0.14 &	0.33 &	   L1641-N	\\ 
26 &	209.870 &	-19.758 &	  4 &	1.00 &	1.5 &	0.14 &	0.06 &	0.46 &	   L1641-N	\\ 
27 &	209.945 &	-19.640 &	  8 &	1.00 &	2.0 &	0.30 &	0.11 &	0.30 &	   L1641-N	\\ 
28 &	210.017 &	-19.659 &	  4 &	1.00 &	1.8 &	0.06 &	0.02 &	0.75 &	   L1641-N	\\ 
29 &	210.047 &	-19.613 &	  5 &	1.00 &	1.3 &	0.18 &	0.07 &	0.55 &	   L1641-N	\\ 
30 &	210.369 &	-19.768 &	  4 &	1.00 &	1.2 &	0.09 &	0.04 &	0.38 &	      V380	\\ 
31 &	210.425 &	-19.716 &	  4 &	1.00 &	0.8 &	0.21 &	0.09 &	0.25 &	      V380	\\ 
32 &	211.011 &	-19.370 &	  4 &	1.00 &	1.4 &	0.11 &	0.05 &	0.63 &	   L1641-C	\\ 
33 &	211.515 &	-19.236 &	  4 &	1.00 &	1.3 &	0.39 &	0.17 &	0.48 &	   L1641-C	\\ 
34 &	212.496 &	-19.016 &	  5 &	1.00 &	1.6 &	0.05 &	0.02 &	0.47 &	   L1641-S	\\ 
35 &	212.579 &	-19.011 &	  5 &	1.00 &	1.9 &	0.14 &	0.05 &	0.34 &	   L1641-S	\\ 
36 &	212.582 &	-18.980 &	  4 &	1.00 &	1.2 &	0.22 &	0.10 &	0.33 &	   L1641-S	\\ 
37 &	212.739 &	-19.053 &	  4 &	1.00 &	1.3 &	0.18 &	0.09 &	0.45 &	   L1641-S	\\ 
\hline 
\end{tabular} 
\tablefoot{(1) String center; (2) Fraction of YSOs (i.e., APOGEE sources with Spitzer and/or XMM counterparts) per string.}
\end{table*} 

\begin{table*} 

\caption{APOGEE strings: individual sources (757 entries, 10 shown; Full table availabe at CDS)} \label{tab:table2} 
\centering 
\begin{tabular}{l c c c c c c} 
\hline 
APOGEE\_ID &	$l$ &		$b$ &	V$_{LSR}$ &	eV$_{LSR}$ &	String$^{(1)}$ &	 YSO?$^{(2)}$  \\ 
&	(deg) &	(deg) &	(km s$^{-1}$) &	(km s$^{-1}$) &	(ID)	 \\  
\hline 
  2M05333429-0444180 &	208.191 &	-19.466 &	10.021 &	0.024 &	  1 &	  Y	\\ 
  2M05340203-0444536 &	208.256 &	-19.368 &	10.021 &	0.145 &	  1 &	  Y	\\ 
  2M05342518-0445109 &	208.307 &	-19.285 &	9.879 &	0.038 &	  1 &	  Y	\\ 
  2M05344863-0447499 &	208.396 &	-19.218 &	9.876 &	0.098 &	  1 &	  Y	\\ 
  2M05343772-0448577 &	208.392 &	-19.267 &	12.314 &	0.053 &	  2 &	  Y	\\ 
  2M05334127-0449257 &	208.286 &	-19.479 &	12.403 &	0.031 &	  2 &	  N	\\ 
  2M05341138-0451229 &	208.377 &	-19.383 &	12.404 &	0.107 &	  2 &	  Y	\\ 
  2M05342224-0457410 &	208.498 &	-19.390 &	12.372 &	0.034 &	  2 &	  N	\\ 
  2M05342978-0451477 &	208.420 &	-19.318 &	12.350 &	0.062 &	  2 &	  Y	\\ 
  2M05344061-0443315 &	208.312 &	-19.215 &	12.318 &	0.084 &	  2 &	  Y	\\
   ... &	...  &	...  &	...  &	...  &	...  &	 	  ... 	\\ 
  \hline 
\end{tabular} 
\tablefoot{(1) Sring ID (see Table~\ref{tab:table1}); (2) Source with Spitzer and/or XMM counterpart?.} 
\end{table*} 

\end{document}